\documentclass[12pt,draftclsnofoot,onecolumn,peerreview,journal]{IEEEtran}
\usepackage{graphicx}
\usepackage{verbatim,cite,flafter,caption,amsmath,amssymb}
\usepackage{mathrsfs}

\makeatletter
\def\fps@figure{hbp}
\makeatother
\begin{document}

\bibliographystyle{ieeetran}
\title{Fast Reliability-based Algorithm of Finding Minimum-weight Codewords for LDPC Codes}
\author{Guangwen Li, Guangzeng Feng}
\maketitle
\begin{abstract}
 Despite the NP hardness of  acquiring minimum distance $d_m$ for linear codes theoretically,
 in this paper we propose one experimental method of finding minimum-weight codewords, the weight of which is equal to $d_m$
 for LDPC codes. One existing syndrome decoding method, called
serial belief propagation (BP) with  ordered statistic decoding (OSD), is adapted to serve our purpose.
We hold the conjecture that among many candidate error patterns in OSD reprocessing, modulo 2 addition of the
lightest error pattern with one of the left error patterns may generate a light codeword. When the decoding syndrome changes to all-zero state,
the lightest error pattern reduces to all-zero, the lightest non-zero error pattern is a valid codeword to update lightest codeword list.
 Given sufficient codewords sending, the survived lightest codewords are likely
to be the target.
Compared with existing techniques, our method demonstrates its efficiency in the simulation of
several interested LDPC codes.
\end{abstract}
\IEEEpeerreviewmaketitle
\section{Introduction}
low-density parity-check (LDPC) codes, as one class
of linear codes, has gained great interest since its rediscovery
by Mackay et al.\cite{ADM:Mac99}, the success is largely due
to presence of belief propagation (BP) decoding achieving near
Shannon-limit performance. In some application, such
as designing or estimating decoding performance by union bound
(UB) in high signal noise ratio (SNR) region, it is desirable to
know the asymptote of UB in advance. However it has been proved in
\cite{Vardy}\cite{Dumer1} that even the minimum distance $d_{m}$
of linear code could not be obtained in polynomial time unless P
= NP. Consequently, minimum-weight codewords, the weight of which
is $d_{m}$ for linear codes, could not be identified in polynomial time either.
The lack of information about minimum-weight
codewords which contribute the most to UB, therefore leads to loose or
inaccurate estimation of UB.  For LDPC codes with medium to long
length, the challenge is for one thing, UB is a useful tool to
analyze its near maximum likelihood decoding (MLD) performance in
the region where Monte Carlo simulation is
unreachable. For another, there exists few candidates among existing
techniques to discern minimum-weight codewords quickly and
reliably with limited computational resource.

Many methods have been proposed to estimate $d_m$ of linear codes.
In \cite{stern1988mfc} and \cite{canteaut1998naf}, probabilistic
algorithms were put forward in finding minimum-weight codewords in
any linear code of medium size, but the computational complexity
will rise speedily with increase of block length. In
\cite{Berrou2002}, one error impulse (EI) method, based on the ability of soft-in decoder,
showed that the
maximum magnitude of the EI, which could be barely corrected  by
the decoder, is directly related to $d_m$ of the linear code.
To tackle $d_m$ of LDPC codes, \cite{Hu2004_1} proposed a randomized
algorithm called nearest nonzero codewords search (NNCS). In this
method, minimal but sufficient noise is purposely imposed on the
all-zero codeword sent, then tentative BP soft information of
each iteration is sent to reliability-based algorithm
\cite{ADM:fossorier2001} for reprocessing, it is expected that the
lightest candidate codewords obtained are exactly the
minimum-weight codewords after trial of all noise patterns. In \cite{Daneshgaran2005}, modification
and extension of \cite{Hu2004_1} was made by employing  EI  method twice in a
two-level search. \cite{richter:fss} introduced an method based on
\cite{Daneshgaran2005} to find small stopping sets in the
bipartite graph of LDPC codes, wherein minimum-weight codeword is
regarded as one special stopping set, it is less complex and works
well for irregular LDPC codes. \cite{hirotomo2005pcm} developed the idea of \cite{stern1988mfc} by applying it for LDPC codes,
one advantage of it is that relations among number of iterations required, number of codewords with weight $w$ and
the probability of codewords with weight $w$ being found in one iteration are described in formulas, which could be utilized
to compare and verify results of various algorithms.

In this paper,  we propose one modification of \cite{ADM:fossorier2001} to acquire
 minimum-weight codewords of LDPC codes
experimentally. All-zero codeword is sent through
AWGN channel with standard variance $\sigma$ being appropriately
set, without EI imposed. Then syndrome decoding based on bit reliability,
in conjunction with standard BP serially , is adapted to generate candidate codewords, with the
lightest ones being recorded. The recorded codewords are likely to be our answer assume that sufficient
 codewords are transmitted.

The remainder of this paper is organized as follows. Section
\makeatletter \@Roman{2} \makeatother details the adaptation of reliability-based syndrome decoding to
obtain minimum-weight codewords.
Simulation result is discussed in
Section \makeatletter \@Roman{3} \makeatother and Section
\makeatletter \@Roman{4} \makeatother concludes our work.
\section{Adaptation of reliability-based syndrome decoding}
\subsection{Implementation of the algorithm}
In \cite{ADM:fossorier2001}\cite{fossorier1995sdd}, the
reliability-based reprocessing, called ordered statistic decoding
(OSD), is involved with most reliable basis (MRB) from columns of
generator matrix $G$. In \cite{fossorier1998rbs},
reliability-based syndrome decoding for linear block codes showed
that the least reliable basis (LRB) of parity check matrix $H$ and
MRB of $G$ are dual of each other, and syndrome decoding  has equivalent error performance to its counterpart in
\cite{fossorier1995sdd}. Considering LDPC codes has sparse $H$ but
dense $G$, we prefer framework of syndrome decoding which is related
with $H$. The merit is that when
Gaussian elimination of $H$ is solicited during reprocessing, the characteristic of sparseness
makes it easier to reduce $H$ instead of $G$ into systematic form
in terms of computational complexity.

Assume binary $(N,K)$ LDPC code with length $N$ and dimension $K$,
then parity check matrix is of the form $H_{M\times{N}}$, where
$M=N-K$ is number of check sums.  BPSK modulation maps
codeword $\overline{c}=[c_1,c_2,\ldots,c_N]$ into
$\overline{x}=[x_1,x_2,\ldots,x_N]$ with $x_i=2c_i-1$, $i\in [1,N]$.
After it is transmitted through AWGN memoryless
channel, we get corrupted sequence
$\overline{y}=[y_1,y_2,\ldots,y_N]$ at receiver, where
$y_i=x_i+z_i$, $z_i$ is independent Gaussian random variable
$\mathscr{N}(0,\sigma^2)$. Hence initial LLR of $i$th bit $v_i$,  is known as
$$
l^0_i=\ln(\frac{p(y_i|c_i=1)}{p(y_i|c_i=0)})=\frac{2y_i}{\sigma^2},
\hspace{3mm} i \in [1,N]
$$
For order-$p$ OSD, it could correct decoding error of standard BP
with at most $p$ erroneous bits in its information set. Naturally
one key point of OSD is  how to define bit reliability reasonably,
since the definition will have impact on which bits are selected
as information set, thus leading to different OSD
performance. We will adopt bit reliability definition of \cite{jiangm2007}, where the reliability $r_i$ of bit $v_i$
is defined as
\begin{equation}
\label{equ:reliability_definition}
r_i=\left|\sum_{j=0}^{j=I_m}\alpha^{I_m-j}\hspace{1mm}l_i^{j}\right|,\hspace{3mm} i \in [1,N]
\end{equation}
where $I_m$ is maximum iteration of BP decoding, $l_i^j$ is LLR of $i$th bit after $j$th iteration and  $\alpha=1$
is assumed in this paper for convenience. As we will see, simulation result in next section justifies the definition of (\ref{equ:reliability_definition}).
The incentive of employing OSD reprocessing to find minimum-weight codewords is based on following conjecture. That is,
 for nonzero decoding syndrome, modulo $2$ addition of two error patterns with small support size will have
higher probability of being one minimum-weight codeword than that with large support size. For the special case of all-zero decoding syndrome, the non-zero error pattern with smallest weight,
or say codeword in such scenario,  has some probability to be one candidate of minimum-weight codewords.

Based on existing
literature\cite{fossorier1995sdd,fossorier1998rbs,ADM:fossorier2001},
the adapted serial BP-OSD to acquire minimum-weight codewords proceeds as follows.
\begin{enumerate}
\item
For the AWGN channel with specified $\sigma$, totally $L_{c}$ codewords are transmitted to receiver.
\item
OSD reprocessing is invoked after $I_m$th iteration of standard BP decoding.
\item
Without losing too much generality, suppose matrix $H$ to be full rank. Permutation
$\lambda_1$ sorts  each bit $e_i, i \in [1,N]$ of error pattern
$\bar{e}$ in ascending order of reliability, and changes $H$ into
$H_1$ by columns reordering. Permutation $\lambda_2$ on $H_1$ is
to ensure the leftmost $M$ columns of resultant $H_2$ to be
independent, thus forming LRB, and the other bit indices
constitute information set. Accordingly original error pattern is converted
into $\bar{e_2}=\lambda_2(\lambda_1(\bar{e}))$.
\item
Apply elementary row operations on both $H_2$ and syndrome
$\bar{s}$ of $I_m$th iteration, so that $H_2$ is transformed into
systematic form. That is
\begin{equation}
\label{equ:matrix_represent} H_2\bar{e_2}=[H_2^1\hspace{3pt}
H_2^2][\bar{e_2^1} \hspace{3pt}
\bar{e_2^2}]^{'}=\bar{s}\Rightarrow
H_2^1\bar{e_2^1}+H_2^2\bar{e_2^2}=\bar{s} \\
\Rightarrow
\bar{e_2^1}=H_2^{1^{-1}}H_2^2\bar{e_2^2}+H_2^{1^{-1}}\bar{s}
\end{equation}
\item For order-$p$ OSD reprocessing,
there are combinations of  $\sum_{i=0}^{i=p}\binom{K}{i}$
candidate error patterns to be reprocessed.
Specifically, for each $\bar{e_2^2}$ in
(\ref{equ:matrix_represent}), assign $1$ to at most $p$ positions
of it, with other positions being zero. Then  $\bar{e_2^1}$
obtained from (\ref{equ:matrix_represent}), in combination with
$\bar{e_2^2}$, forms a distinct error pattern $e_2=[\bar{e_2^1}
\hspace{3pt} \bar{e_2^2}]$.
\item
After reordering those error patterns in ascending order of Hamming weight,
for nonzero decoding syndrome, modulo $2$ addition of the first error pattern with each of the left error patterns will generate
one valid codeword, record the one(s) with lightest weight; For all-zero decoding syndrome, the nonzero lightest
codeword(s) could be identified instantly. Then update minimum-weight codeword list with above result.
\item
Return to step $2$ to continue another decoding attempt till decoding of  all $L_c$ codewords is checked. Lastly, the survived
minimum-weight codewords will represent as the estimation for the interested LDPC code.
\end{enumerate}
\subsection{Selection of the key parameters}
Noticeably, for our approach, simulation shows appropriate setting of $\sigma$ and $I_m$  will save lots of computational complexity.
Suppose standard BP implementation of \cite{neal_software}, all-zero codeword is transmitted, then $I_m=a$
is determined if  codeword bit $v_i$ satisfies
\begin{equation*}
l_i^{a}\nrightarrow -\infty,\hspace{3mm} \forall\hspace{1mm}i \in [1,N]\text{\newline}
\end{equation*}
\begin{equation*}
l_i^{a+1}\rightarrow -\infty,\hspace{3mm} \exists  \hspace{1mm}i \in [1,N]
\end{equation*}
Evidently the sense of $-\infty$ is coherent with BP implementation. For the choice of $\sigma$,
two factors have to be considered. First it should be small as possible so that the corrupted sequence in signal space
is near the origin in terms of Euclidean distance, ensuring that it has high probability to  be decoded correctly  by standard BP or OSD reprocessing.
More importantly, $\sigma$ should be large enough so that BP decoding with sufficient iterations are solicited, which manifests
strength of definition (\ref{equ:reliability_definition}) for reprocessing. So one desirable scenario is that
corrupted sequence is rarely decoded successfully at $0$th iterationy of standard BP, but shows near MLD performance after $I_m$ iteration.
Although above guidelines could give roughly selection of $I_m$ and $\sigma$, the optimal
values of them still resort to simulation result.

\section{Simulation Result and Discussion}
To make computational complexity manageable on notebook AMD Athlon
$1800+$ with $252$M RAM, $p=2$ is set for order-$p$ OSD
reprocessing. For $C_0: 96.33.964 (96,48)$, $C_1: 495.62.3.2915$
$(495,433)$, $C_2: 252.252.3.252 (504,252)$ and $C_3:504.504.3.504
(1008,504)$ in \cite{mackay_database}, simulation settings and
result are listed in \ref{table:1}. Running time refers to the
processing time of order-2 OSD for all $L_c$ codewords. The last
column denotes when the minimum-weight codeword and its
multiplicity is identified as earliest actually. Because of
randomness of AWGN channel, the data listed should be translated
statistically.
\begin{table}
\centering \caption{\label{table:1} Estimation performance of the
interested LDPC codes}
\begin{tabular}{|c|c|c|c|c|c|c|}
\hline
Code & $\sigma$ & $I_m$&$L_c$ & $(d_m, Multi.)$ & running time(Hour)& earliest $n$th \\
\hline
 $C_0$&0.70&5&100&(6,2)&0.01&4\\
 $C_1$&0.44&4&100&(4,60)&0.08&60\\
 $C_2$&0.70&5&1000&(20,2)&0.8&470\\
 $C_3$&0.75&6&10000&(30,1)&140&2599\\
\hline
\end{tabular}
\end{table}
The minimum-weight codeword and its multiplicity in the
\ref{table:1} conform well to the data exposed in
\cite{hirotomo2005pcm}. It was reported in \cite{hirotomo2005pcm}
that it takes $44$ hours, $37$ hours, and $210$ hours for $C_1,
C_2, C_3$ respectively to obtain low weight distribution on the
powerful microcomputer, our approach focuses on minimum-weight
codewords with far less computation resources. Though it is not so
convincing to declare our method is more efficient than
\cite{hirotomo2005pcm}, the observation is that for
\cite{hirotomo2005pcm}, its complexity of bit operations is with
the form $r*O(N^3)$, where $r$ is maximum iteration of Stern's
algorithm \cite{stern1988mfc}. Likely, the complexity of our
method is with the form $L_c*O(N^3)$, $L_c$ is the amount of
codewords sent out which satisfies $L_c<<r$, say $r=10^7$ in
\cite{hirotomo2005pcm} and $L_c=10^4$ in our method for $C_3$. For
NNCS approach,since Gaussian elimination is called every iteration
during one decoding, its efficiency is far less than ours under
the condition of handling same number of corrupted sequences.

With increase of the LDPC code length, it is demonstrated required
$L_c$ increases too. The reason is that error pattern of long code
has the tendency to reverse bits more than $p=2$ in information
set, which lowers the probability of lightest codewords being dug
out under the condition of $p\leq2$. To compensate for such
performance fading, more codewords sending is expected to hold the
probability. Unfortunately, growth of $L_c$ and code length both
will urge much more computation. For instance, the processing time
of $C_2$ is $0.8$ hour, while that of $C_3$ rises to $140$ hours.
\section{Conclusions}
In this paper, we adapt serial BP-OSD algorithm to find
minimum-weight codewords for LDPC codes. Different from previous
work, our method concentrates on finding minimum-weight codewords
only.  The conjecture we holds is that for syndrome decoding of
OSD reprocessing, it is likely that the modulo 2 addition between
candidate error patterns may generate the lightest codewords,
given sufficient codewords sending. The worth of our method over
existing techniques is that better tradeoff between computational
complexity and performance is achieved, simulation result
justifies our approach with several instances of interested LDPC
codes.

\bibliography{IEEEabrv,output1}
\end{document}